%% file: template.tex
\documentclass[cameraready]{Interspeech}

\title{Acoustic Landmark Detector based on Conformer and HuBERT}

\author[affiliation={1,2}, correspondingauthor]{Mateo}{C\'{a}mara}
\author[affiliation={1}]{Jos\'{e} Luis}{Blanco}
\author[affiliation={3}]{Juan Ignacio}{Godino-Llorente}
\author[affiliation={2}]{Jeung-Yoon}{Choi}
\author[affiliation={2}]{Stefanie}{Shattuck-Hufnagel}

\address{
    $^1$ \blue{Signal Processing Applications Group, Information Processing \& Telecomm. Center, Universidad Polit\'{e}cnica de Madrid, Spain} \\
    $^2$ Speech Communication Group, Research Laboratory of Electronics, Massachusetts Institute of Technology, USA \\
    $^3$ Bioengineering and Optoelectronics Lab., Universidad Polit\'{e}cnica de Madrid, Spain
}

\email{mateo.camara@upm.es, \blue{jl.blanco@upm.es, ignacio.godino@upm.es, jyechoi@mit.edu, sshuf@mit.edu}}


\newcommand{\blue}[1]{#1}

\usepackage{comment}

\begin{document}

\maketitle

\begin{abstract}
Acoustic landmarks (abrupt acoustic changes tied to speech events) offer a linguistically grounded representation for speech analysis. We study automatic landmark detection with Conformer models, evaluating 14 configurations spanning architecture, loss, label representation, feature extractor, and data conditions on 1,839 manually annotated utterances with eight landmark types. We propose Gaussian soft labels with per-class temporal spread ($\sigma$=10–20 ms), improving F1@20 ms by 7.0\% absolute vs.\ hard labels by modeling annotation variability. Frozen HuBERT features perform best without fine-tuning (F1@20 ms=0.77). Stops and fricatives are reliable (F1$>$0.80), while vowels remain challenging (F1$\approx$0.55). On our corpus, our system reaches a 13.8\% Landmark Error Rate (LER). This is not directly comparable to AutoLandmark (31.3\%) or SpeechMark (56.5\%), evaluated on a different corpus and metric. Per-class trends show detectability increases with event abruptness, consistent with Stevens' theory.
\end{abstract}

\section{Introduction}
\label{sec:intro}

In the landmark-based lexical access framework, acoustic landmarks are salient temporal events in the speech signal (often discontinuities or extrema) that mark cues to distinctive features~\cite{stevens2002acoustic}. Eight landmark types are distinguished: vowels~(V), glides~(G), and closure/release pairs for stops~(Sc/Sr), fricatives~(Fc/Fr), and nasals~(Nc/Nr)~\cite{choi_1997}. These landmarks provide a linguistically motivated proxy that bridges raw acoustics and phonological features, with applications in speech recognition~\cite{hasegawa2005landmark}, speech/clinical assessment~\cite{boyce2012speechmark}, and related timing-based analyses~\cite{shattuck-hufnagel_robustness_2007}.

Automatic landmark detection has historically been dominated by signal-processing heuristics. While deep learning has achieved strong performance in related phoneme boundary detection tasks~\cite{franke2016phoneme,kreuk2020selfsupervised}, \blue{landmark detection targets a different, sparser representation: rather than segmenting every phone boundary, it localizes and labels a linguistically defined set of articulatory events.} Existing work leaves open how far current deep models can go in identifying and timing acoustic landmarks from the signal alone.

We address this gap with a Conformer-based~\cite{gulati2020conformer} landmark detection system evaluated across 14~experimental configurations. Our contributions are: (1)~a Gaussian soft-label training strategy with per-class $\sigma$ values (10--20\,ms) that accounts for annotation variability across landmark types. (2)~A systematic comparison of mel spectrograms, wav2vec2~\cite{baevski2020wav2vec}, HuBERT~\cite{hsu2021hubert}, and hybrid features for landmark detection. (3)~Ablation studies covering loss functions, model capacity, and data conditions. and (4)~A per-class analysis linking detection difficulty to the acoustic properties predicted by Stevens' theory~\cite{stevens2002acoustic}.

\section{Related Work}
\label{sec:related}

\subsection{Classical landmark detection.}
Stevens' landmark theory~\cite{stevens2002acoustic, yun2020landmark} provides the acoustic--phonological foundation for this work. Liu~\cite{liu1996landmark} proposed an early automatic landmark detector using energy-band analysis and rule-based classification. Howitt~\cite{howitt2000automatic} improved vowel landmark detection with neural networks applied to formant-related acoustic parameters. The 2004 Johns Hopkins workshop~\cite{hasegawa2005landmark} integrated landmark detection into a recognition pipeline, and Jansen and Niyogi~\cite{jansen2008landmark} introduced hierarchical wavelet features for robust extraction. Recent cue-based landmark work suggests that statistical classifiers such as Gaussian Mixture Models can be effective for landmark detection as well \cite{park_meng2025}. All these systems relied on hand-engineered acoustic features with limited contextual modeling.

\subsection{Deep learning for speech events.}
Neural approaches have significantly advanced the related task of phoneme boundary detection. Franke et~al.~\cite{franke2016phoneme} applied bidirectional LSTMs to TIMIT segmentation, while Kreuk et~al.~\cite{kreuk2020selfsupervised} achieved unsupervised phone segmentation with contrastive learning. Forced alignment tools such as the Montreal Forced Aligner~\cite{mcauliffe2017montreal} typically \blue{use} HMM-based approaches but target phone-level boundaries rather than articulatory landmarks. Sound event detection~\cite{mesaros2016metrics} shares the challenge of temporal localization in audio, including evaluation with tolerance-based metrics analogous to ours.

\subsection{Auto-Landmark.}
Most closely related to our work, Zhang et~al.~\cite{zhang2024autolandmark} recently released Auto-Landmark, a TIMIT-based dataset with manually refined landmark timing (described as the first publicly available resource of this kind), along with an open-source signal-processing toolkit and deep learning baselines. They annotated the TIMIT corpus~\cite{garofolo1993timit} with five landmark types (glottis, burst, sonorant, fricative, voiced fricative) and evaluated with Landmark Error Rate (LER), computed on landmark patterns without considering timing, and (following earlier conventions) without distinguishing onset vs. offset. Their best deep learning model (ConBiMamba) achieved 31.3\% LER, substantially outperforming the signal-processing baseline SpeechMark~\cite{boyce2012speechmark} (56.5\% LER). Our work differs in three key respects: (1)~we use Stevens' eight-type landmark taxonomy rather than their five-type set. (2)~We employ a frame-level Conformer with tolerance-based F1 evaluation that explicitly rewards temporal precision. (3)~We train on a separate corpus of isolated syllables and words rather than TIMIT continuous speech.

\subsection{Self-supervised speech representations.}
We evaluate frozen Self-Supervised Learning (SSL) features for landmark detection. Pre-trained models such as wav2vec\,2.0~\cite{baevski2020wav2vec} and HuBERT~\cite{hsu2021hubert} learn speech representations from unlabeled data, including cues related to manner of articulation, which are directly relevant for landmark identification. \blue{Benchmarks such as SUPERB~\cite{yang2021superb} motivate using SSL representations across speech tasks, though landmark detection is not among them.} 

\subsection{Soft labels and temporal smoothing.}
Our Gaussian soft-label strategy is related to temporal label smoothing methods used in action detection~\cite{yeche2023temporal} and onset detection~\cite{schluter2014improved}, where class boundaries are inherently ambiguous. Unlike uniform label smoothing, our per-class $\sigma$ encodes the expected temporal precision of each landmark type, inspired by the observation that different articulatory events have different acoustic durations and annotation reliabilities~\cite{yun2020landmark}.

\section{Database}
\label{sec:dataset}

\begin{figure}[t]
    \centering
    \includegraphics[width=0.66\columnwidth]{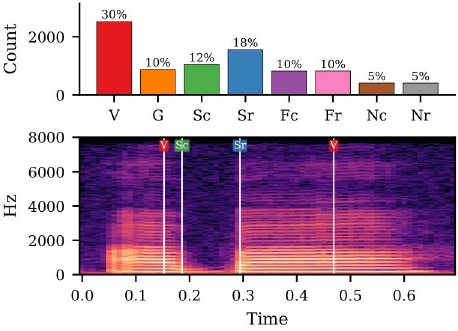}
    \caption{(a) Landmark type distribution. (b) Spectrogram with annotated landmarks for a VCV utterance.}
    \label{fig:dataset}
\end{figure}

Our corpus comprises 1,839 speech recordings (mean duration 0.80\,s) from three speakers (one female, two male), annotated with acoustic landmarks in Praat TextGrid format.\footnote{Audio examples and landmark visualizations: \url{https://mateocamara.github.io/acoustic-landmarks/}} The dataset contains 678 vowel-consonant-vowel (VCV) syllables providing controlled phonetic contexts and 1,161 real English words with diverse phonological structures. Annotations follow taxonomy from ~\cite{choi_1997} with eight landmark types across 8,428 total instances (mean 4.6 per file). Significant class imbalance exists (see Figure~\ref{fig:dataset}a). Audio is resampled to 16\,kHz. The train/test split is 90/10 with speaker stratification across both subsets.

\section{Method}
\label{sec:method}

\subsection{Model Architecture}

Our primary architecture is a Conformer encoder~\cite{gulati2020conformer} with $d_\text{model}$\,=\,256, 12~layers, 4~attention heads, feed-forward dimension 1024, and depthwise convolution kernel size~31. Each block follows the Conformer structure: half-step feed-forward, multi-head self-attention, convolution module, and half-step feed-forward, with layer normalization. The encoder output feeds a linear classification head with dropout~0.3, producing per-frame logits over 9~classes (background + 8~landmarks). The model processes complete utterances non-causally. We also evaluate a \emph{per-category} variant with five separate smaller Conformer models for vowels, glides, stops, fricatives, and nasals.

\subsection{Gaussian Soft Labels}
\label{sec:softlabels}

Standard hard labels assign a single frame per landmark, ignoring the inherent uncertainty in human annotation timing. The soft-label formulation has two effects: it provides a margin around the annotated position during training, and it concentrates the learning signal proportionally to the expected temporal precision of each class. We replace labels with Gaussian-weighted soft label distributions:
\begin{equation}
    y_{t,c} = \exp\left(-\frac{(t - t_c^*)^2}{2\sigma_c^2}\right),
    \label{eq:gaussian}
\end{equation}
where $t_c^*$ is the annotated time and $\sigma_c$ is a per-class temporal spread that controls how much probability mass is distributed around the annotated position during \emph{training}. We set $\sigma$ \blue{as a phonetic prior (not a tuned search),} based on the expected acoustic duration of each event type: $\sigma_V$\,=\,20\,ms for vowels (gradual spectral transitions), $\sigma_G$\,=\,15\,ms for glides, $\sigma_{Fc,Fr}$\,=\,12\,ms for fricatives, and $\sigma_{Sc,Sr,Nc,Nr}$\,=\,10\,ms for stops and nasals (abrupt events). The background probability is $1 - \max_c y_{t,c}$, and the distribution is normalized to sum to~1. Note that these $\sigma$ values govern label smoothing during training and are distinct from the evaluation tolerance windows described in Section~\ref{sec:setup}. 

\subsection{Feature Extraction}

We evaluate four feature types: (1)~80-dimensional log-mel spectrograms with 25\,ms window, 10\,ms hop, and per-utterance normalization. (2)~Frozen wav2vec2-base~\cite{baevski2020wav2vec} (768-dim, 20\,ms frame shift). (3)~Frozen HuBERT-base~\cite{hsu2021hubert} (768-dim, 20\,ms). (4)~A hybrid concatenating mel with wav2vec2 (848-dim, interpolated to 10\,ms alignment).

\subsection{Post-Processing}

Frame-level probability curves are obtained via softmax. For each non-background class, we apply peak detection with minimum height~0.5, minimum inter-peak distance of 5~frames, and minimum prominence~0.2. Each detected peak yields a landmark prediction at the corresponding time with the peak probability as confidence.

\section{Experimental Setup}
\label{sec:setup}

All models are trained with AdamW ($\text{lr}$\,=\,$10^{-4}$, weight decay 0.01), cosine annealing with warm restarts, automatic mixed precision, gradient clipping at~1.0, and early stopping (patience~15). Data augmentation includes time stretching ($\pm$10\%), pitch shifting ($\pm$2 semitones), additive noise (SNR 20--40\,dB), and SpecAugment~\cite{park2019specaugment}. We additionally leverage a synthetic dataset of auto-labeled landmarks to add out-of-domain training data~\cite{pink_trombone_english_landmarks}. We compare weighted cross-entropy (inverse-frequency class weights), focal loss~\cite{lin2017focal} ($\gamma$\,=\,2), and unweighted cross-entropy.

Predictions are evaluated with tolerance-based matching: a prediction is a true positive if a ground-truth landmark of the same class exists within a tolerance window, using greedy closest-first matching. We compute F1 across 10--50\,ms tolerances and take the plateau of the F1 curve as the primary tolerance, identifying 20\,ms as the primary metric (F1@20\,ms), with F1@30\,ms as secondary. We evaluate 14~configurations spanning feature extractors, loss functions, label strategies, model capacities, and data conditions.

\section{Results}
\label{sec:results}

\subsection{Overall Performance}

\input{tables/table2_main_results}

\begin{figure}[t]
    \centering
    \includegraphics[width=0.68\columnwidth]{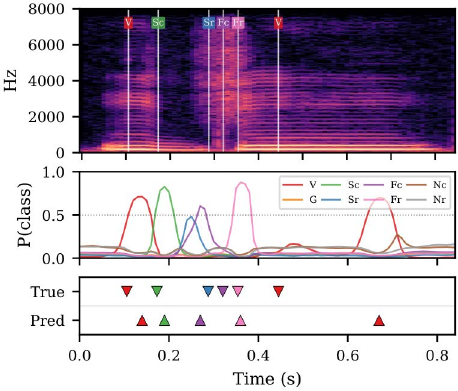}
    \caption{Example prediction from the baseline model for utterance ``iydjiy'': (a) spectrogram with ground-truth landmarks, (b) class probability curves, (c) predicted vs.\ true landmark.}
    \label{fig:example}
\end{figure}

Table~\ref{tab:results} presents per-class F1@20\,ms for four feature configurations alongside a hard-label reference. Frozen HuBERT features achieve the best overall performance (F1\,=\,0.77), followed by the hybrid mel+wav2vec2 model (0.76), the mel baseline (0.74), and wav2vec2 (0.70). HuBERT and hybrid models both achieve F1@30\,ms\,=\,0.84. Wav2vec2 underperforms the mel baseline despite its richer representations, likely because its coarser 20\,ms frame shift reduces temporal precision.
\subsection{Effect of Gaussian Soft Labels}

\begin{figure}[t]
    \centering
    \includegraphics[width=0.56\columnwidth]{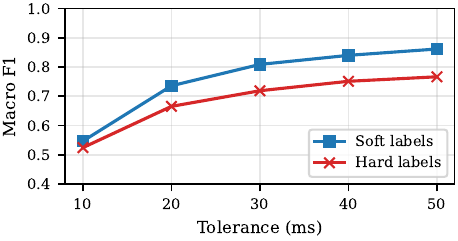}
    \caption{F1 as a function of tolerance window: the gap between soft and hard labels widens from 10 to 50\,ms, indicating better localization rather than broader detection.}
    \label{fig:softlabel}
\end{figure}

Gaussian soft labels provide the single largest improvement across all design choices evaluated. Hard labels degrade F1 by $-$0.070 relative to the soft-label baseline, with catastrophic impact on vowels (F1 dropping from 0.54 to 0.18. See Table~\ref{tab:results}). This confirms that vowels benefit most from the temporal margin that soft labels provide.

Figure~\ref{fig:softlabel} shows that the soft-label advantage grows with tolerance: the gap between soft and hard labels widens from 0.022 at 10\,ms to 0.070 at 20\,ms. The soft-label model improves by $+$0.189 from 10 to 20\,ms tolerance, while the hard-label model gains only $+$0.141, indicating that soft-label models produce more concentrated predictions near the true position.
\subsection{Per-Class Analysis}

Table~\ref{tab:results} reveals systematic patterns in detection difficulty. Stop and fricative releases and stop closures are easiest (Sr, Fr, Sc; F1\,$>$\,0.80), while vowels (V) and nasal releases (Nr) are consistently the hardest, reflecting their gradual acoustic manifestations. This hierarchy aligns with Stevens' prediction~\cite{stevens2002acoustic} that abrupt articulatory gestures produce more distinctive acoustic signatures.

Cross-domain analysis reveals that the HuBERT model generalizes well across subsets (VCV: 0.78, words: 0.76), while the mel baseline shows a stark asymmetry (VCV: 0.59, words: 0.78), suggesting that SSL features provide more domain-invariant representations.

\subsection{Comparison with Prior Work}

\input{tables/table4_prior_work}

Table~\ref{tab:prior_work} contextualizes our results against prior landmark detection systems. Direct comparison is difficult because earlier work used different corpora, landmark inventories, and evaluation protocols. Earlier systems reported partial results on smaller landmark inventories (Table~\ref{tab:prior_work}): rule-based energy-band analysis~\cite{liu1996landmark}, formant-based neural networks~\cite{howitt2000automatic}, and wavelet features with SVM classifiers~\cite{jansen2008landmark}. Our system detects all eight landmark types with tolerance-based F1 evaluation, achieving 0.77~F1@20\,ms (a stricter metric that requires correct classification and temporal localization).

\subsection{Cross-corpus comparison with Auto-Landmark.}
To assess generalization beyond our corpus, we evaluated our models on the 1,680 TIMIT~\cite{garofolo1993timit} test files used by Auto-Landmark~\cite{zhang2024autolandmark} (without TIMIT training data). The two systems use different landmark taxonomies: Auto-Landmark defines five types (glottis, burst, sonorant, fricative, voiced fricative), while we detect eight Stevens-type landmarks. We established correspondences for three pairs: fricatives (their f$\pm$ $\leftrightarrow$ our Fc/Fr), stops (b$\pm$ $\leftrightarrow$ Sc/Sr), and nasals (s$\pm$ $\leftrightarrow$ Nc/Nr).

On the Auto-Landmark LER metric (ignoring timing), their ConBiMamba reaches 31.3\% LER, whereas our HuBERT model yields 63.0\% on the subset with three clean correspondences. This gap is expected because LER emphasizes sequence ordering and is optimized by their CTC setup. Importantly, the zero-shot results are not uniform across categories: fricatives transfer reaches 0.39 LER, stops 0.53, and nasals 0.73, indicating only a partially transferable landmark structure across taxonomies. 

\section{Ablation Studies}
\label{sec:ablations}

\input{tables/table3_ablation}

Table~\ref{tab:ablation} summarizes ablation results relative to the mel baseline. The \blue{feature-extractor comparisons appear in the main results (Table~\ref{tab:results}).}

\subsection{Loss Function and Class Weights}

Focal loss~\cite{lin2017focal} ($-$0.048) optimizes frame-level accuracy at the cost of peak sharpness needed for landmark extraction. Removing class weights has a moderate effect ($-$0.027), indicating that inverse-frequency weighting is important but not critical.

\subsection{Model Capacity and Architecture}

Both small ($d_\text{model}$\,=\,128, 6~layers; $-$0.016) and large ($d_\text{model}$\,=\,512, 16~layers; $-$0.014) variants perform near baseline, indicating that the default 12-layer model already has sufficient capacity for this dataset size. The per-category architecture ($-$0.096) fragments training data across five models, eliminating shared learning across landmark types.

\subsection{Data Conditions}

Training on VCV syllables only ($-$0.101) causes the largest degradation, showing that phonetically diverse data is essential. Words-only training (+0.032) slightly improves over the mixed baseline. Data augmentation (+0.002) and synthetic pretraining (+0.000) provide negligible benefit, suggesting the current corpus sufficiently covers the acoustic variation in the test set.

\section{Conclusions}
\label{sec:conclusions}

We presented a Conformer-based system for acoustic landmark detection, achieving F1@20\,ms of~0.77 with frozen HuBERT features. \blue{On our corpus, our model reaches a 13.8\% LER and covers more landmark types (8 vs.\ 5--6). This is not directly comparable to Auto-Landmark's 31.3\% on TIMIT (different corpus, inventory, and metric).}
Gaussian soft labels with per-class $\sigma$ provide the largest single improvement ($+$0.070 F1), highlighting label representation as a key design choice for temporally localized speech events.
Limitations include a small corpus (1{,}839 files, 3 speakers), a single train/test split, and limited zero-shot transfer to TIMIT for stop-landmark detections. \blue{We do not claim broad generalization to other speaking styles or recording conditions.} Future work includes larger corpora, fine-tuning SSL encoders \blue{(and WavLM or Whisper)}, vowel-specific strategies, and evaluation on continuous, spontaneous speech.

\section{Acknowledgements}
This work was supported by the Ministry of Economy and Competitiveness of Spain under grant PID2021-128469OB-I00, and by the ``Ayuda Econ\'{o}mica Para Personal Investigador Postdoctoral 2025'' of the Fundaci\'{o}n Santander. This work was also supported by a Grant of the MISTI MIT Global Experiences.

\section{Generative AI Use Disclosure}
Generative AI tools have only been used for editing/polishing (not for writing any major part). The authors reviewed all the text.

\bibliographystyle{IEEEtran}
\bibliography{mybib}

\end{document}

%% file: tables/table2_main_results.tex
\begin{table*}[t]
\caption{Main results: per-class F1 at 20\,ms tolerance. Best per-column values in bold. Hard-label reference uses mel features for comparison with soft-label systems.}
\label{tab:results}
\centering
\small
\begin{tabular}{lcccccccccc}
\toprule
\textbf{System} & \textbf{V} & \textbf{G} & \textbf{Sc} & \textbf{Sr} & \textbf{Fc} & \textbf{Fr} & \textbf{Nc} & \textbf{Nr} & \textbf{F1@20} & \textbf{F1@30} \\
\midrule
Baseline (mel) & \textbf{0.54} & \textbf{0.70} & \textbf{0.85} & 0.91 & 0.74 & 0.82 & 0.76 & 0.57 & 0.74 & 0.81 \\
wav2vec2 & 0.43 & 0.52 & 0.83 & 0.87 & 0.73 & 0.79 & 0.89 & 0.54 & 0.70 & 0.78 \\
HuBERT & 0.53 & 0.69 & 0.83 & \textbf{0.93} & \textbf{0.78} & 0.89 & 0.86 & \textbf{0.62} & \textbf{0.77} & \textbf{0.84} \\
Hybrid (mel+wav2vec2) & 0.53 & 0.61 & 0.81 & \textbf{0.93} & 0.72 & \textbf{0.90} & \textbf{0.93} & \textbf{0.62} & 0.76 & \textbf{0.84} \\
\midrule
Hard labels (reference) & 0.18 & 0.63 & 0.81 & 0.85 & 0.71 & 0.89 & 0.73 & 0.54 & 0.67 & 0.72 \\
\bottomrule
\end{tabular}
\end{table*}

%% file: tables/table4_prior_work.tex
\begin{table}[t]
\caption{Comparison with prior landmark detection systems. \blue{Systems differ in corpus, landmark inventory, and metric; values are not directly comparable.}}
\label{tab:prior_work}
\centering
\small
\resizebox{\columnwidth}{!}{%
\begin{tabular}{lccl}
\toprule
\textbf{System} & \textbf{Types} & \textbf{Metric} & \textbf{Approach} \\
\midrule
Liu~\cite{liu1996landmark} & 6 & 73\% acc. & Rule-based \\
Howitt~\cite{howitt2000automatic} & V only & 90\% rec. & NN + formants \\
Jansen~\cite{jansen2008landmark} & 6 & 80\% acc. & Wavelet + SVM \\
SpeechMark~\cite{boyce2012speechmark} & 5 & 56.5 LER & Rule-based \\
Auto-LM~\cite{zhang2024autolandmark} & 5 & 31.3 LER & ConBiMamba \\
\midrule
Ours  & 8 & 0.77 F1 | 13.8 LER & Conformer + HuBERT \\
\bottomrule
\end{tabular}
}
\end{table}

%% file: tables/table3_ablation.tex
\begin{table}[t]
\caption{Ablation study grouped by factor, relative to mel baseline (F1@20ms = 0.736).}
\label{tab:ablation}
\centering
\small
\resizebox{\columnwidth}{!}{%
\begin{tabular}{llcc}
\toprule
\textbf{Category} & \textbf{Variant} & \textbf{F1@20ms} & \textbf{$\Delta$} \\
\midrule
Loss & Focal loss & 0.688 & -0.048 \\
 & No class weights & 0.709 & -0.027 \\
\midrule
Capacity & Small model & 0.720 & -0.016 \\
 & Large model & 0.722 & -0.014 \\
\midrule
Architecture & Per-category & 0.640 & -0.096 \\
\midrule
Data & VCV only & 0.635 & -0.101 \\
 & Words only & 0.768 & +0.032 \\
\midrule
Augmentation & No augmentation & 0.738 & +0.002 \\
\midrule
Transfer & Synth. pretrain & 0.736 & +0.000 \\
\bottomrule
\end{tabular}
}
\end{table}